\documentclass[aps,pre,showpacs,twocolumn]{revtex4-1}
\usepackage{graphicx}

\begin{document}
\title{Spectral properties of microwave graphs with local absorption}

\author{Markus Allgaier}
\affiliation{Fachbereich Physik der Philipps-Universit\"{a}t Marburg, D-35032 Marburg, Germany}
\author{Stefan Gehler}
\affiliation{Fachbereich Physik der Philipps-Universit\"{a}t Marburg, D-35032 Marburg, Germany}
\author{Sonja Barkhofen}
\affiliation{Fachbereich Physik der Philipps-Universit\"{a}t Marburg, D-35032 Marburg, Germany}
\author{Ulrich Kuhl}
\email{ulrich.kuhl@unice.fr}
\affiliation{LPMC, CNRS UMR 7336, Universit\'{e} de Nice Sophia-Antipolis, F-06108 Nice, France}
\affiliation{Fachbereich Physik der Philipps-Universit\"{a}t Marburg, D-35032 Marburg, Germany}
\author{H.-J. St\"{o}ckmann}
\email{stoeckmann@physik.uni-marburg.de}
\affiliation{Fachbereich Physik der Philipps-Universit\"{a}t Marburg, D-35032 Marburg, Germany}

\date{\today}

\begin{abstract}
The influence of absorption on the spectra of microwave graphs has been studied experimentally. The microwave networks were made up of coaxial cables and T junctions. First, absorption was introduced by attaching a 50\,$\Omega$ load to an additional vertex for graphs with and without time-reversal symmetry. The resulting level-spacing distributions were compared with a generalization of the Wigner surmise in the presence of open channels proposed recently by Poli \emph{et al.}\ [Phys. Rev. Lett. {\bf 108}, 174101 (2012)]. Good agreement was found using an effective coupling parameter. Second, absorption was introduced along one individual bond via a variable microwave attenuator, and the influence of absorption on the length spectrum was studied. The peak heights in the length spectra corresponding to orbits avoiding the absorber were found to be independent of the attenuation, whereas, the heights of the peaks belonging to orbits passing the absorber once or twice showed the expected decrease with increasing attenuation.
\end{abstract}


\maketitle

\section{Introduction}

Quantum graphs, consisting of connected networks of bonds and vertices, are an ideal playing ground to study questions coming from quantum chaos and random matrix theory (RMT). For example the trace formula expressing the spectra of graphs in terms of periodic orbits is exact, in contrast to its equivalent for quantum-chaotic systems, the Gutzwiller trace formula \cite{gut90}. Moreover, a classification of the periodic orbits in terms of a symbolic alphabet is straight forward for graphs. Details can be found in the paper by Kottos and Smilansky \cite{kot99a}.

In closed quantum graphs the main interest focused on the statistical properties of the spectra. Most studies in this respect were motivated by the famous conjecture by Bohigas, Giannoni, and Schmit (BGS) stating that the universal features of the spectra of chaotic systems can be described by RMT \cite{boh84b} (see also Ref.~\cite{cas80}). Using supersymmetry techniques Gnutzmann and Altland \cite{gnu04b} succeeded to prove the BGS conjecture for the two-point correlation function for graphs with incommensurate bond lengths, provided the underlying classical dynamics is chaotic. Their result was recently generalized for all correlation functions by Pluha{\v r} and Weidenm\"{u}ller \cite{arXplu13}. Graphs without time-reversal invariance (TRI) should, hence, share their universal properties with those of the Gaussian unitary ensemble (GUE) and systems with TRI and no half-integer spin with those of the Gaussian orthogonal ensemble (GOE). Experimentally, this conjecture was tested in microwave graphs for the nearest-neighbor level-spacing distribution. Good agreement with the Wigner distributions predicted by RMT for graphs with and without TRI was found \cite{hul04}.

For closed systems, Wigner distributions are known to be good approximations for the nearest-neighbor level-spacing distribution in chaotic systems, which were obtained using a two by two Hamiltonian (see e.\,g. Refs.~\cite{stoe98, haa01b}). New features come into play if the graph is gradually opened. For a weak opening, individual resonances can still be observed, and they are only slightly shifted into the complex plane. For such a situation, Poli \emph{et\,al.}~\cite{pol12} proposed a generalization of the Wigner distribution being exact for $2\times 2$ matrices and one attached channel. Additionally, they found numerically that this surmise is a good approximation for arbitrary matrix ranks and arbitrary channel numbers if the channel coupling strength is used as an effective parameter.

If the graph is opened even more, up to the point where the widths of the resonances become comparable with or even larger than the mean level spacings, individual resonances can no longer be resolved and  scattering theory comes in. On the experimental side, there are investigations on the Wigner reaction matrix \cite{hul05a}, the elastic enhancement factor \cite{law10}, and graphs with iso-spectral scattering properties  \cite{hul12}. The theoretical studies of the scattering properties of quantum graphs again started with a paper by Kottos and Smilansky \cite{kot03} and are still an active part of research \cite{mar13,gnu13,arXgut,plu13a,plu13b,arXplu13}.

Starting with a description of the experiment in the next section, in Sec.~III, an experimental test of the level-spacing distributions of Poli \emph{et\,al.}\ in open microwave graphs is presented. In Sec.~IV, the spectra of graphs with a variable attenuator along one of the bonds are discussed in terms of periodic orbits.

\section{Experiment}
\label{sec:experiment}

\begin{figure}
  \centering
  \raisebox{3cm}{(a)}
	\includegraphics[width=0.8\columnwidth]{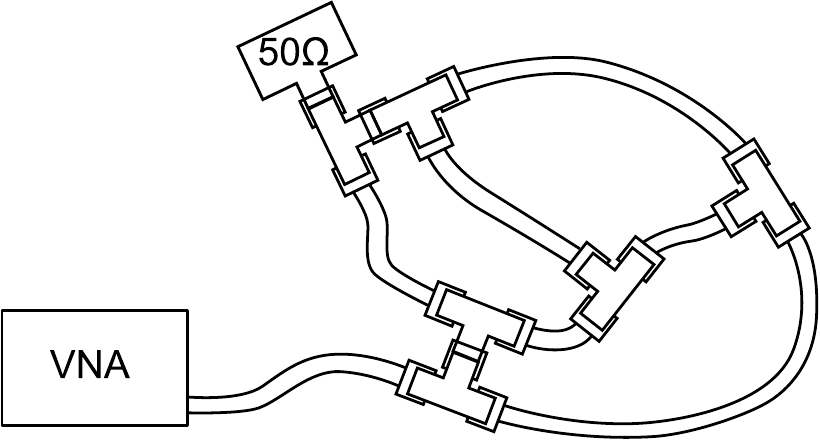}\\
  \raisebox{3cm}{(b)}
	\includegraphics[width=0.8\columnwidth]{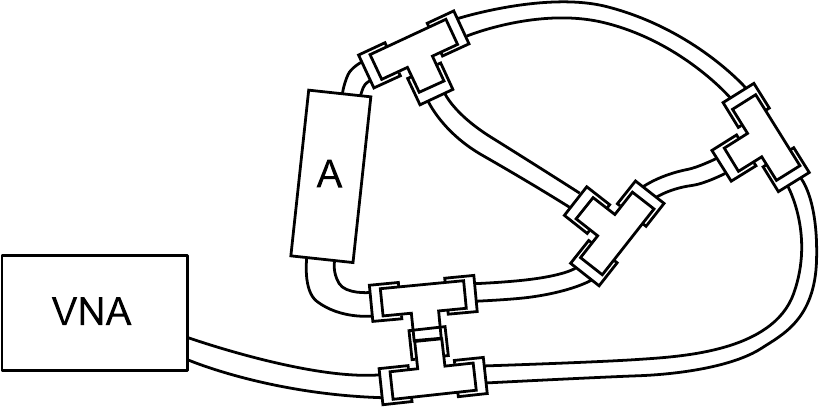}\\
  \caption{\label{fig:setup}
Microwave graphs: (a) Graph with an additional open channel which is closed by a 50\,$\Omega$ terminator. (b) Graph where inside a single bond, an attenuator is attached. The first graph (a) has been used for the study of level-spacing statistics, and the second one (b) has been used for the investigation of the length spectrum.}
\end{figure}

In the experiments we used microwave networks to simulate quantum graphs similarly as they had been used by Hul et al.~\cite{hul04,hul05a,law10} (see Fig.~\ref{fig:setup}). The bonds are formed by Huber~\&~Suhner EZ-141 coaxial semi-rigid cables with SMA connectors, coupled by T junctions at the vertices.  The inner and outer radii of the cables are 0.45 and 1.45\,mm, respectively, hence, below 34.8 GHz only the lowest TEM mode is propagating (see Sec.~8.8 of Ref.~\cite{jac98}). Reflection measurements were performed with an Agilent 8720ES vector network analyzer (VNA) coupled to one of the vertices (see Fig.~\ref{fig:setup}).

The optical lengths of all cables were determined experimentally in a separate measurement. The scattering matrix of the T junction (such as all scattering matrices) may be written as $S=VS_0W$, where $V$, $W$ are unitary and $S_0$ is orthogonal \cite{bee97}. $V$, $W$  take care of the phase shifts on the leads within the connector and may be absorbed in a redefinition of the bond lengths. Without loss of generality, we may, hence, assume that $S$ is orthogonal. The T junctions were found to be symmetric with respect to the three ports, corresponding to a scattering matrix
\begin{equation}\label{eq:scatt}
    S=\frac{1}{3}\left(\begin{array}{ccc}
                         -1 & 2 & 2 \\
                         2 & -1 & 2 \\
                         2 & 2 & -1
                       \end{array}\right)\,,
\end{equation}
up to a sign, the only possibility for symmetric connectors being in accordance with orthogonality. This means nothing but current conservation. The sign has been chosen in accordance with Ref.~\cite{kot99a}. In quantum transport measurements a different sign convention is applied \cite{bee97}. The quantum mechanical analog of the microwave network is, hence, a quantum graph with von Neumann boundary conditions at the vertices.

The cables and T junctions were assembled to form connected tetrahedral microwave graphs. We introduce absorption locally in two different ways. The first way is an additional channel to the environment realized by a 50\,$\Omega$ terminator connected to one of the vertices [see Fig.~\ref{fig:setup}(a)]. The second way is a variable attenuation in one of the bonds [see Fig.~\ref{fig:setup}(b)].

Using the additional channel, we investigate the level-spacing statistics, e.g., the distance between the real parts of the eigenvalues. To improve statistics, the results for different graphs and different positions of the 50\,$\Omega$ terminator were superimposed. For a part of these measurements, one of the T junctions was replaced by an Aerotek I70-1FFF microwave circulator to break the time-reversal symmetry. A circulator corresponds to a T junction with unidirectional properties allowing transport only from port 1 to port 2, port 2 to port 3, and port 3 to port 1 but not in the opposite direction.

A HP~8494H microwave attenuator was introduced at one of the bonds allowing for attenuations between 0 and 11\,dB in steps of 1\,dB. With this setup,  the effect of local absorption on the complex eigenvalues has been studied, especially its relation to the length spectrum.

\begin{figure}
  \raisebox{5cm}{(a)}\hspace*{-2em}\includegraphics[width=.95\columnwidth]{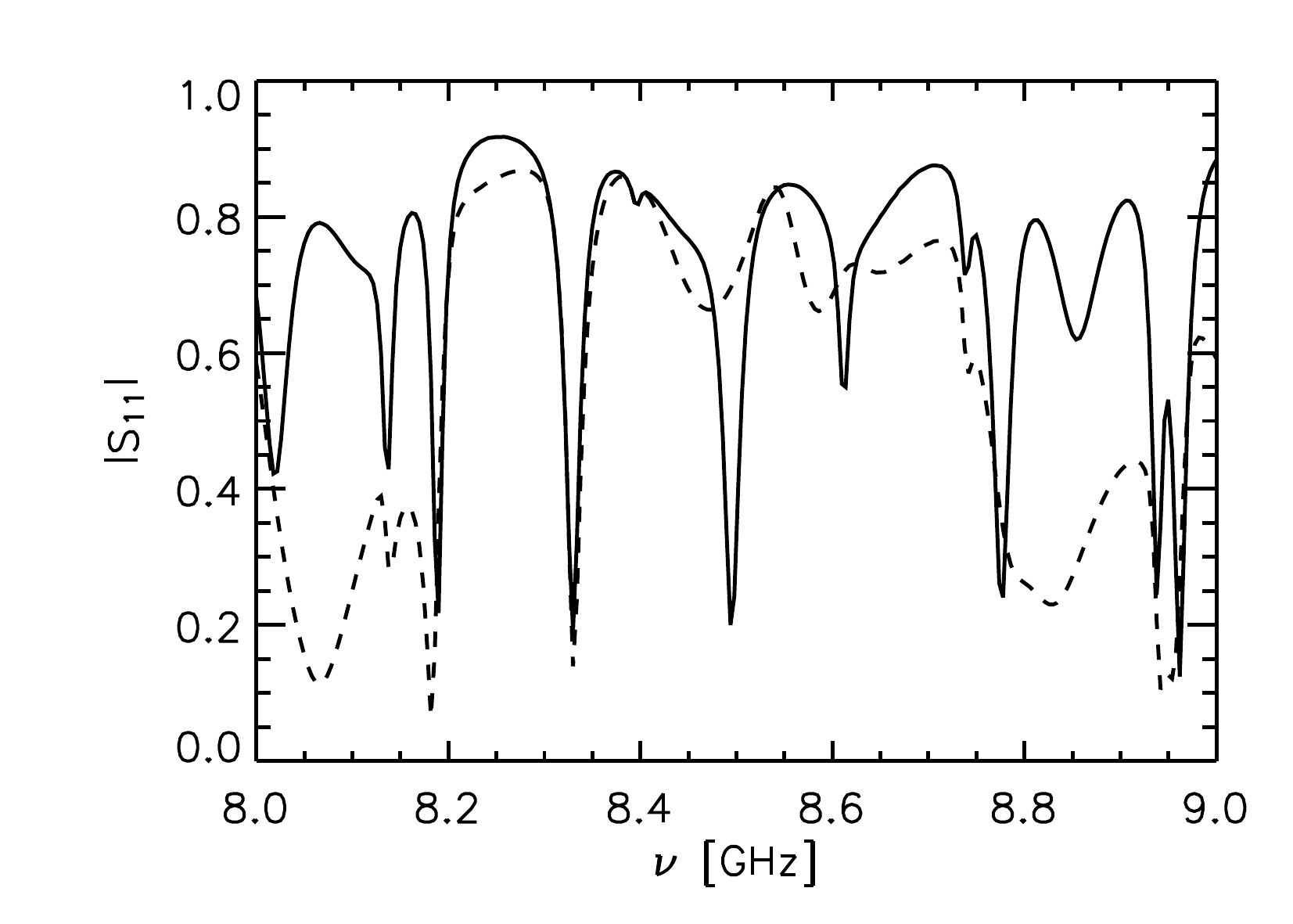}\\
  \raisebox{5cm}{(b)}\hspace*{-2em}\includegraphics[width=.95\columnwidth]{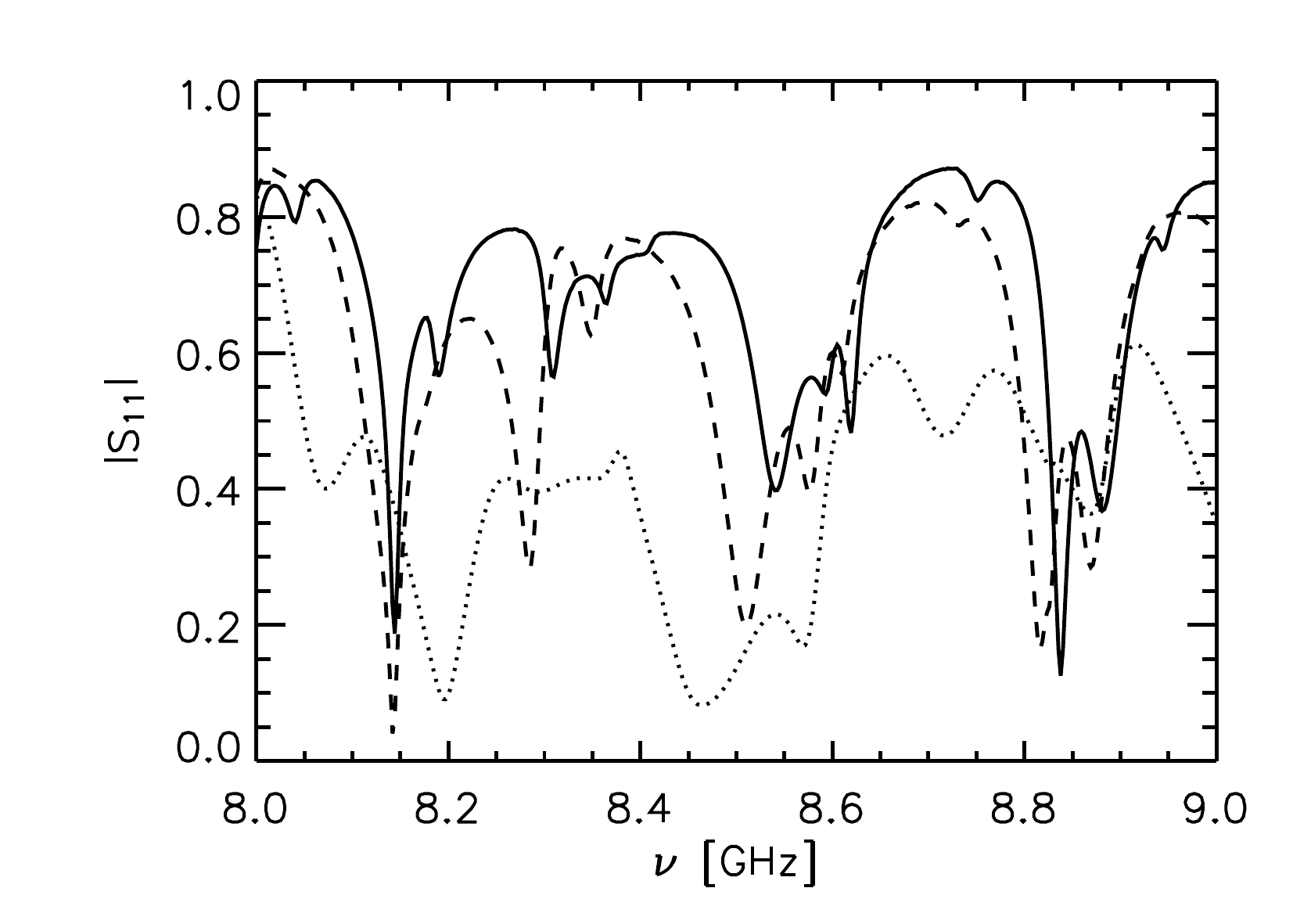}\\
  \caption{\label{fig:spectrum}
  Typical spectra obtained for the two types of graphs shown in Fig.~\ref{fig:setup}. (a) shows spectra for the GOE graph with one (solid line) and two (one additional 50\,$\Omega$ load, dashed line) attached channels. (b) shows spectra for three different values of additional attenuation [0\,dB (solid), 2\,dB (dashed), and 11\,dB (dotted)] on one bond.}
\end{figure}

All experiments were performed in the single mode regime. The measurements on GOE graphs were performed from 4 to 18\,GHz. Due to the limitation of the circulators operating frequency range from 6 to 12\,GHz only this frequency range was investigated for the GUE spacing distribution. For the length spectra investigation we used the operating frequency range of the attenuator (0-18\,GHz). Figure~\ref{fig:spectrum} shows typical reflection spectra for the two types of  graphs.

\section{Level-spacing distributions}
\label{sec:poli}

For many years, RMT has been known to give an excellent description of the universal properties of the spectra of chaotic systems. The quantity most often studied in the past was the level-spacing distribution of neighbored eigenvalues \cite{stoe99,haa01b,kah62,ber84b,bar00b}. RMT yields analytic expressions for the ensemble average of this quantity, where the average is taken over the GOE for the system with time-reversal symmetry and no spin 1/2 and GUE for the system with broken time-reversal symmetry \cite{meh91}. For $2\times 2$ matrices, the ensemble averaged level-spacing distributions are described by the famous Wigner distributions. They deviate from the exact expressions in the large $N$ limit, where $N$ is the rank of the matrix, only by several percent \cite{haa01b}, a deviation usually too small to be detected in experiments. This is why the Wigner distribution, in particular, for the GOE case, enjoys a particular popularity.

This was the motivation for Poli \emph{et\,al.}~\cite{pol12} to extend the Wigner surmise to open systems. They obtained an exact expression for the distribution ${\cal P}_\beta(s)$ for the spacings $s$ of the real parts of the eigenvalues of $2\times 2$ matrices with one attached open channel, where $1/\eta$ corresponds to the channel coupling strength. In the case of time-reversal systems, corresponding to the GOE ($\beta=1$), it reads
\begin{eqnarray}\label{eq:poligoe}
  {\cal P}_1(s)&=&\frac{A\eta}{16}e^{-(A/2)s^2} \int_0^{\infty}dx \frac{1}{\sqrt{s^2+\frac{x^2}{4}}}\,e^{-(A/16)x^2-x\eta/2}\nonumber\\
  &&\times\left[ (8s^2+x^2)I_0 \left(\frac{Ax^2}{16}\right)+x^2I_1\left(\frac{Ax^2}{16}\right)\right]
\end{eqnarray}
where $I_k(z)$ is a modified Bessel function. For systems without time reversal invariance, corresponding to the GUE ($\beta=2$), they found
\begin{equation}\label{eq:poligue}
  {\cal P}_2(s)=e^{-(A/2)s^2}\sqrt{\frac{A}{2\pi}}\eta\left(E(A,\eta)s^2+\frac{2}{\eta^2}-\frac{E(A,\eta)}{A}\right),
\end{equation}
where
\begin{equation}\label{eq:Eaeta}
    E(A,\eta)=e^{\eta^2/2A}E_1\left(\frac{\eta^2}{2A}\right)
\end{equation}
with the exponential integral $E_1(z)=\int_z^\infty \textrm{d}x\frac{e^{-x}}{x}$. In both cases, $A$ fixes the mean level spacing. Poli \emph{et\,al.}\ found that the channel coupling strength $1/\eta$ can be used as an effective parameter to get a good description of the numerical results for all matrix ranks $N$ and the number of coupled channels $M$ tested in the paper.

In this part of the paper we present experimental tests of the Poli \emph{et\,al.}\ distributions in the graphs. For the TRI systems (GOE), we realized seven different microwave graphs without and six microwave graphs with one additional 50\,$\Omega$ terminator. The graphs consist of cables with lengths ranging from 366 to 600\,cm and 533 to 600\,cm, respectively. In the case of the non-TRI systems (GUE), we realized eight realizations without and nine with one additional 50\,$\Omega$ terminator. In this case, the cables length range from 394 to 574\,cm and 431 to 624\,cm, respectively. Adding a second additional 50\,$\Omega$ terminator lead to such a strong damping that a reliable extraction of the eigenvalues was no longer possible. Thus, we stick to one additional channel.

\begin{figure}
  \raisebox{5cm}{(a)}\hspace*{-2em}\includegraphics[width=.95\columnwidth]{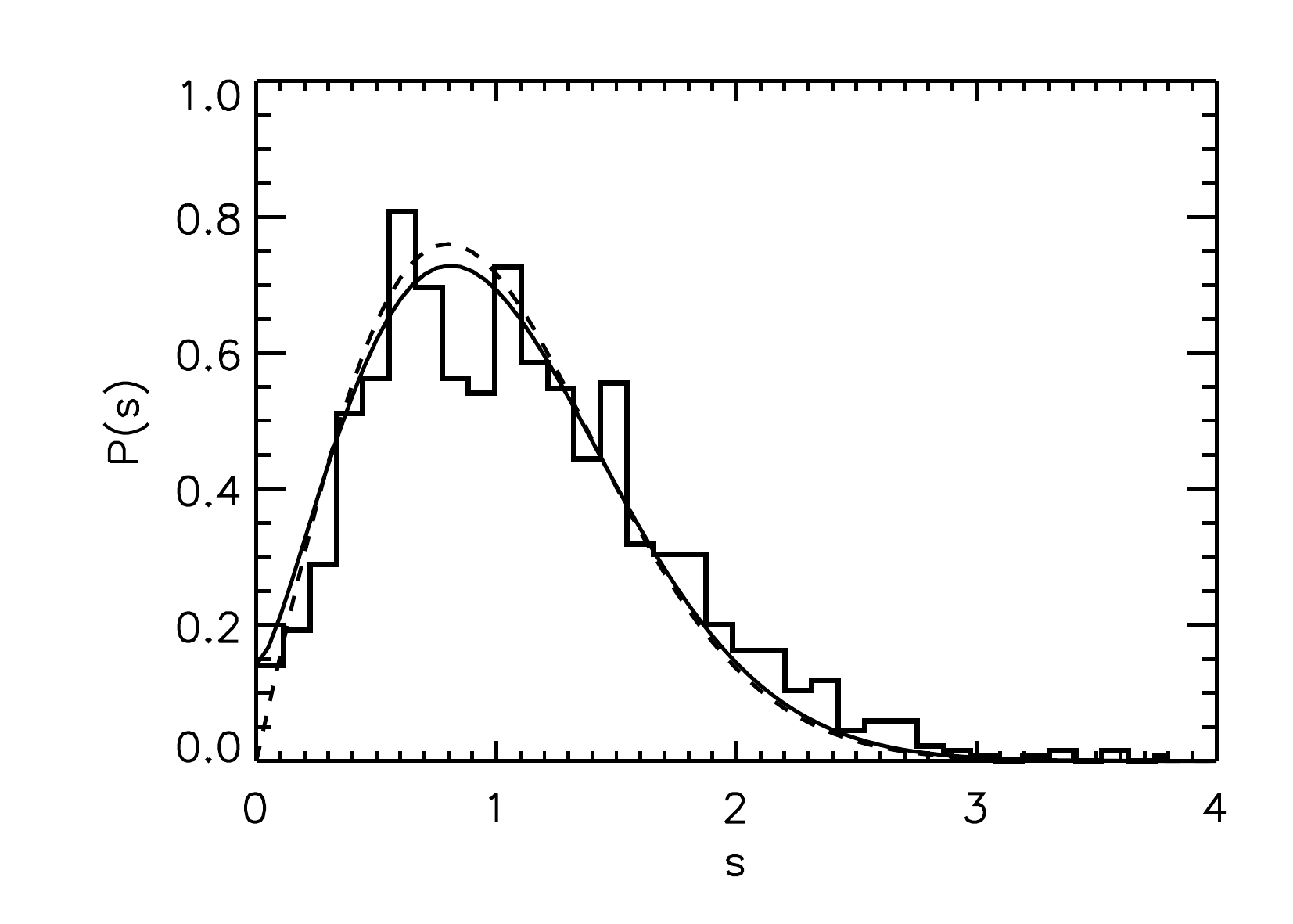}\\
  \raisebox{5cm}{(b)}\hspace*{-2em}\includegraphics[width=.95\columnwidth]{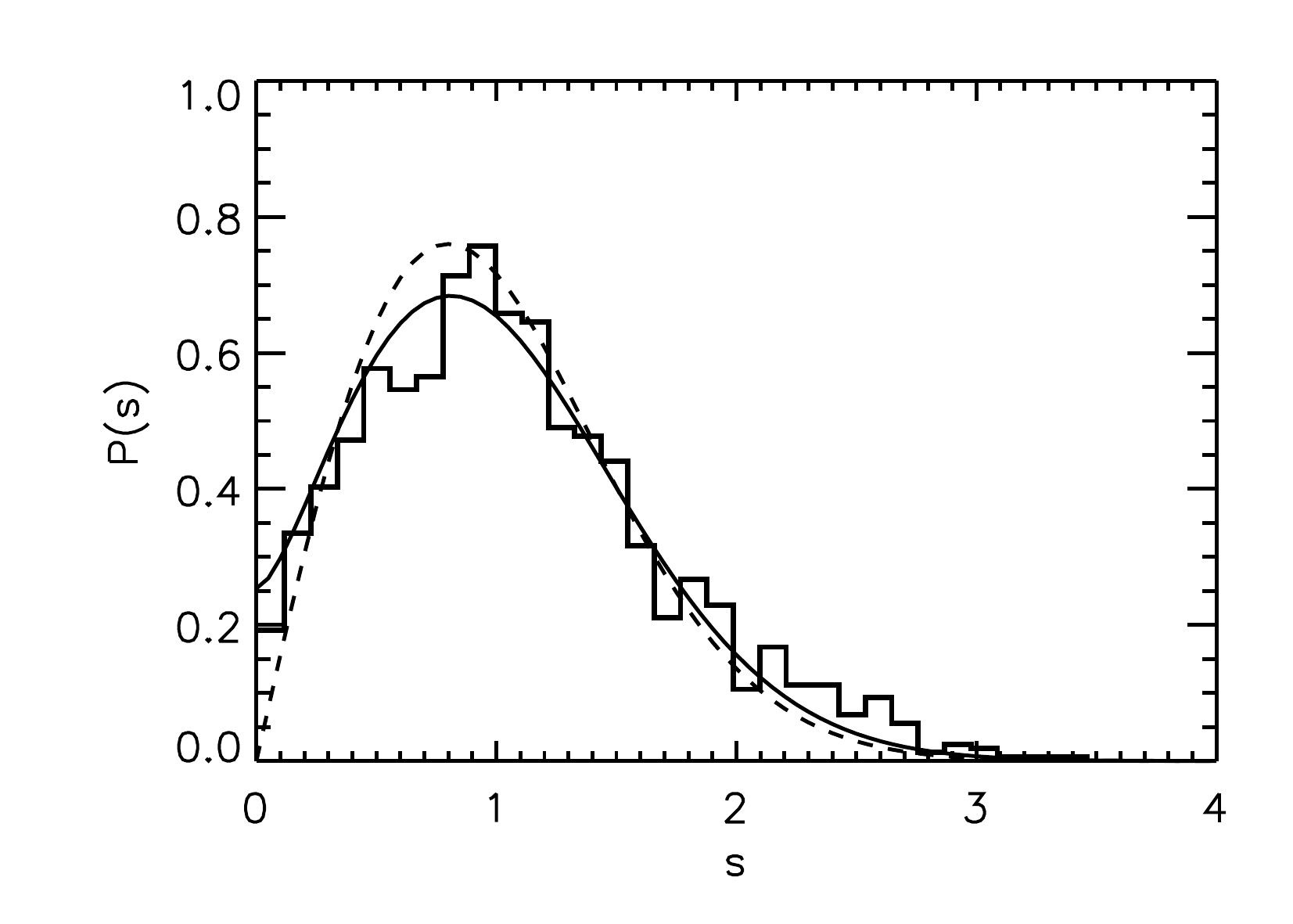}\\
  \caption{\label{fig:GOEpoli}
  Level-spacing distribution for the real parts of the eigenvalues for a graph with time-reversal invariance with (a) the measuring channel only, and with (b) an additional 50\,$\Omega$ terminator at one of the vertices. The dashed lines correspond to the Wigner prediction for GOE systems, and the solid line corresponds to the Poli \emph{et\,al.}\ distribution where the effective channel coupling strength $1/\eta_{\rm eff}$ has been obtained by a fit. In (a), the distribution was obtained from 1228 resonances and $1/\eta_{\rm eff}$ = 0.19 and in (b) the distribution was obtained from 1468 resonances and $1/\eta_{\rm eff}$ = 0.32.}
\end{figure}

\begin{figure}
  \raisebox{5cm}{(a)}\hspace*{-2em}\includegraphics[width=.95\columnwidth]{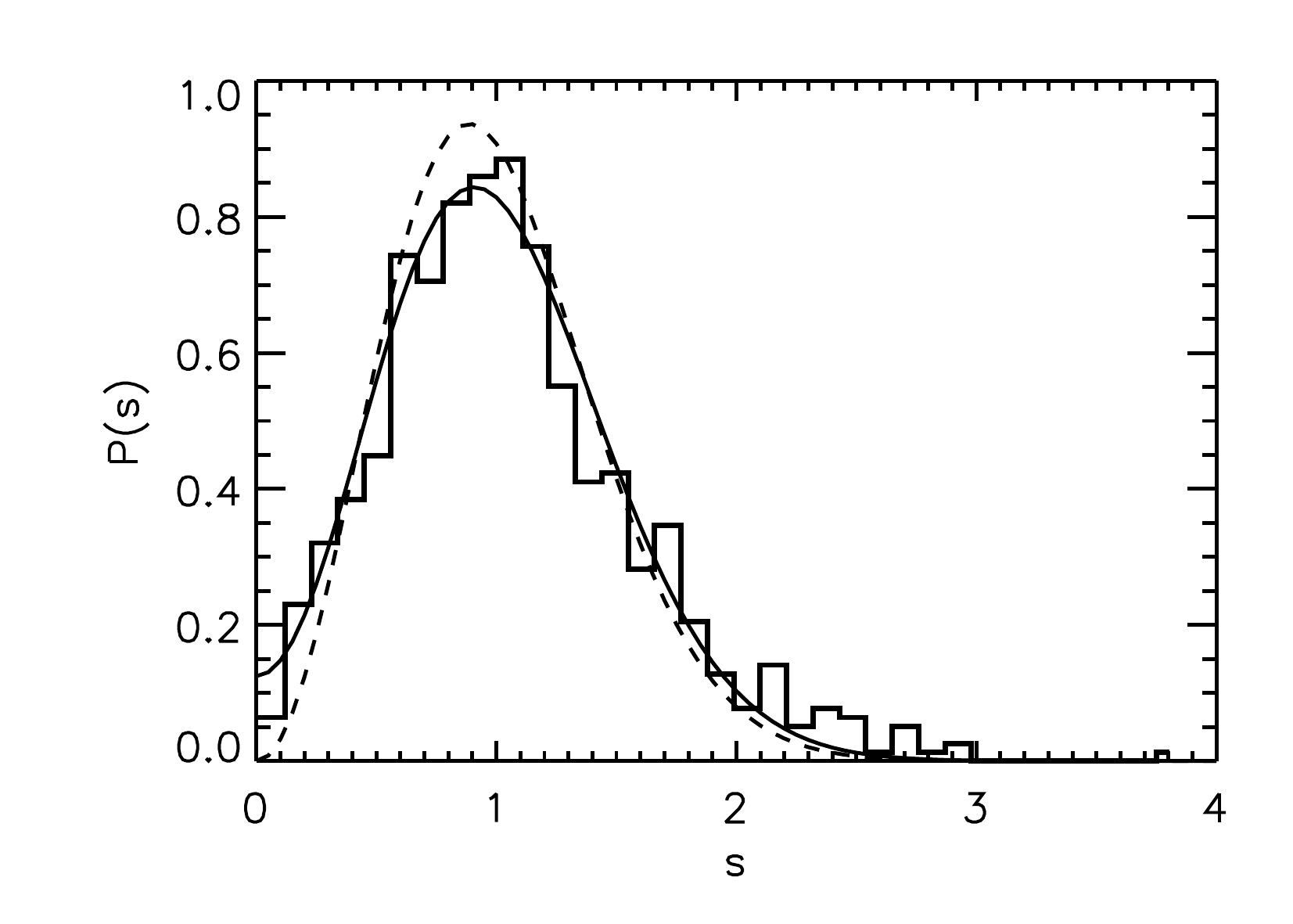}\\
  \raisebox{5cm}{(b)}\hspace*{-2em}\includegraphics[width=.95\columnwidth]{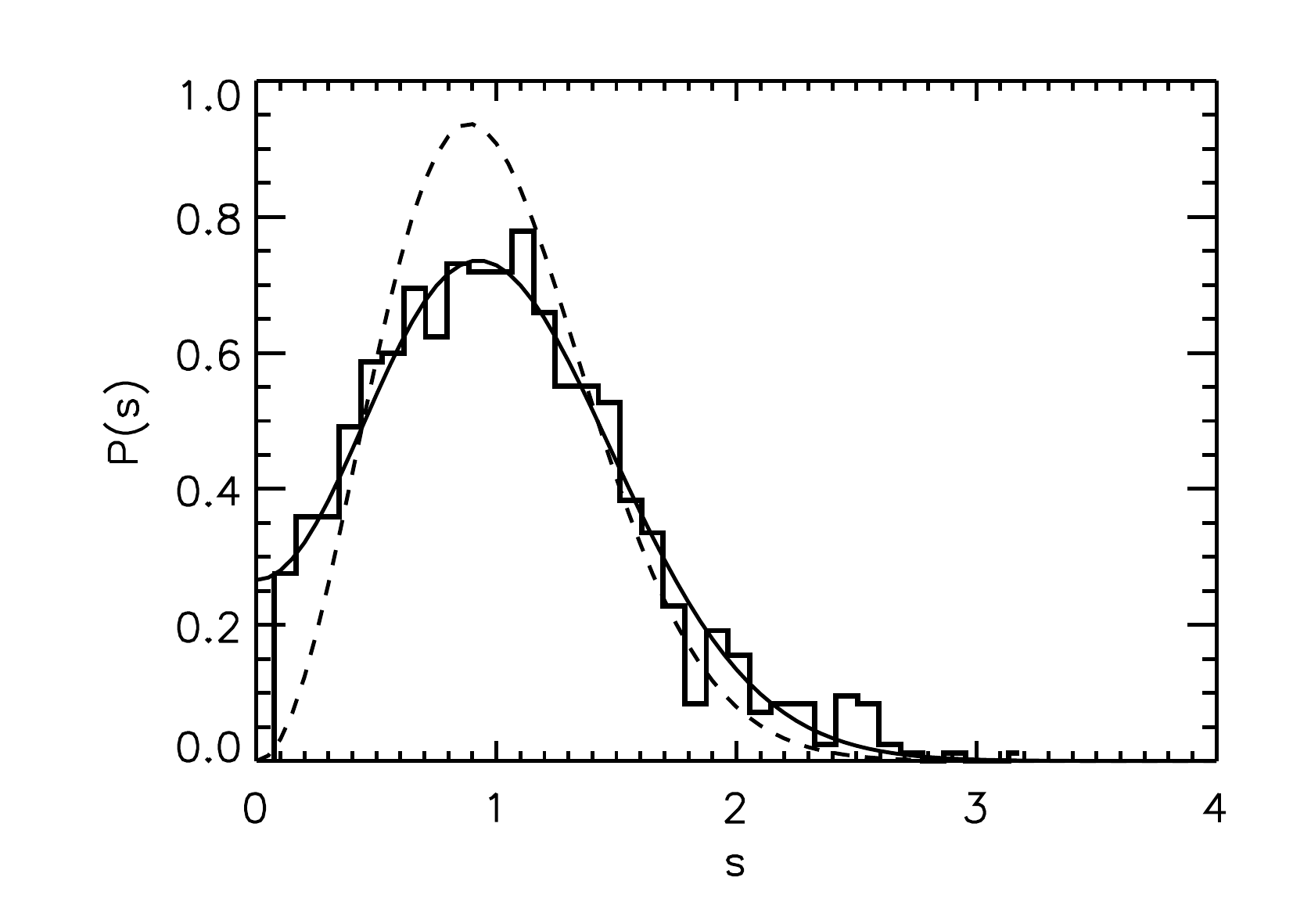}\\
  \caption{\label{fig:GUEpoli}
  Level-spacing distribution for the real parts of the eigenvalues for a graph with broken time-reversal symmetry (GUE) (a) without and (b) with an additional 50\,$\Omega$ load at one of the vertices. The dashed lines correspond to the Wigner prediction for GUE systems, and solid line corresponds to the Poli \emph{et\,al.}\ distribution, where the effective channel coupling strength $1/\eta_{\rm eff}$ has been obtained by a fit. In (a), the distribution was obtained from 710 resonances and $1/\eta_{\rm eff}$ = 0.17 and in (b), from 929 resonances and a $1/\eta_{\rm eff}$ = 0.31.}
\end{figure}

For all spectra presented in this paper, real and imaginary parts of the eigenvalues have been obtained from complex multi-Lorenz fits to the reflection signal $S_{11}$. The results are collected in Figs.~\ref{fig:GOEpoli} and \ref{fig:GUEpoli} for the graphs with and without TRI, respectively. All spectra have been unfolded to a mean level spacing $\Delta$ of one by using the Weyl formula for closed graphs ($\Delta=\pi/L$). The figures show the level-spacing distributions with and without an additional 50\,$\Omega$ load. This corresponds to one and two open channels, respectively, since the attached VNA is equivalent to another 50\,$\Omega$ load.  Microwave networks are based on a 50\,$\Omega$ technology, meaning that, for 50\,$\Omega$ terminators, there is an ideal impedance matching with no reflection at the end. The solid lines correspond to the Poli\emph{et\,al.}\ distributions where the channel coupling strength has been adjusted by a fit. For the fit, we discarded the first bin, as very small spacings cannot be resolved, and discarded values for spacings $s$ larger than 2. Due to the additional channel, the hole in the original Wigner distribution for small distances now is partially filled. This does not mean that there no longer is a level repulsion. In the complex plane, the eigenvalues still repel each other, but for their real parts, corresponding to a projection onto the real axis, this is no longer true. In the previous paper by Hul \emph{et\,al.}~\cite{hul04}, this has not been observed as they coupled the graphs not via a T junction but via a six junction, thereby reducing the coupling to the VNA considerably.

The resulting coupling constants are shown in Table~\ref{tab:eta} where the errors correspond to one standard deviation of the fit. In billiards, the explicit calculation of the coupling strengths is difficult (see, e.\,g.,~Ref.~\cite{bar05a}), but for graphs, it does not pose problems since the scattering properties of the junctions are known. The calculation of the scattering matrix for graphs follows exactly the same route as for billiards, see Sec.~6.1.2 of Ref.~\cite{stoe98}. One obtains a value of $1/\eta= 1/(2\pi)$ for the coupling constant for one channel in units of the mean level spacing. This is in good agreement with the experimentally found values, both for the GOE and for the GUE. For two channels, the Poli \emph{et\,al.}\ distribution yields only an effective coupling constant, mimicking two coupled channels by a single one. Provided the line widths are small compared to the mean level spacing, i.\,e., in the so-called Breit-Wigner approximation, the effective coupling constant for the two channels should be just twice the coupling constant for a single channel, i.\,e., $1/\eta_{\rm eff} = 1/\pi$. Again, a good agreement is found with the experimental values, in fact, a bit too good since, in the case of two coupled channels, the Breit-Wigner approximation, strictly speaking, is no longer justified.

In principle, the coupling strength of the channels could be determined independently from a reflection measurement using $T_{a}=1-|\langle S_{aa}\rangle|^2=4 \textrm{Re}(1/\eta)/|1+1/\eta|^2$ where the average is taken over the ensemble \cite{ver85a,koeb10}. For a sufficient statistics, however, one would have to also average over frequency being unreliable because of global phase changes with frequency.

\begin{table}
\begin{ruledtabular}
\begin{tabular}{|l|c|c|}
& One channel& Two channels\\\hline
GOE& 0.19 $\pm$ 0.05& 0.32 $\pm$ 0.05\\
GUE& 0.17 $\pm$ 0.05& 0.31 $\pm$ 0.05\\\hline
Theory& $1/(2\pi)$=0.159&$1/\pi$=0.318\\
\end{tabular}
\end{ruledtabular}
\caption{\label{tab:eta}
Effective channel coupling constants $1/\eta_{\rm eff}$, obtained by fitting the Poli \emph{et\,al.}\ distributions for the GOE and GUE to the experimental level-spacing distributions. The theoretical predictions hold for open channels ideally coupled via a T junction (see text for details).}
\end{table}

\section{Periodic orbits}
\label{sec:orbits}

\begin{figure}
  \includegraphics[width=.95\columnwidth]{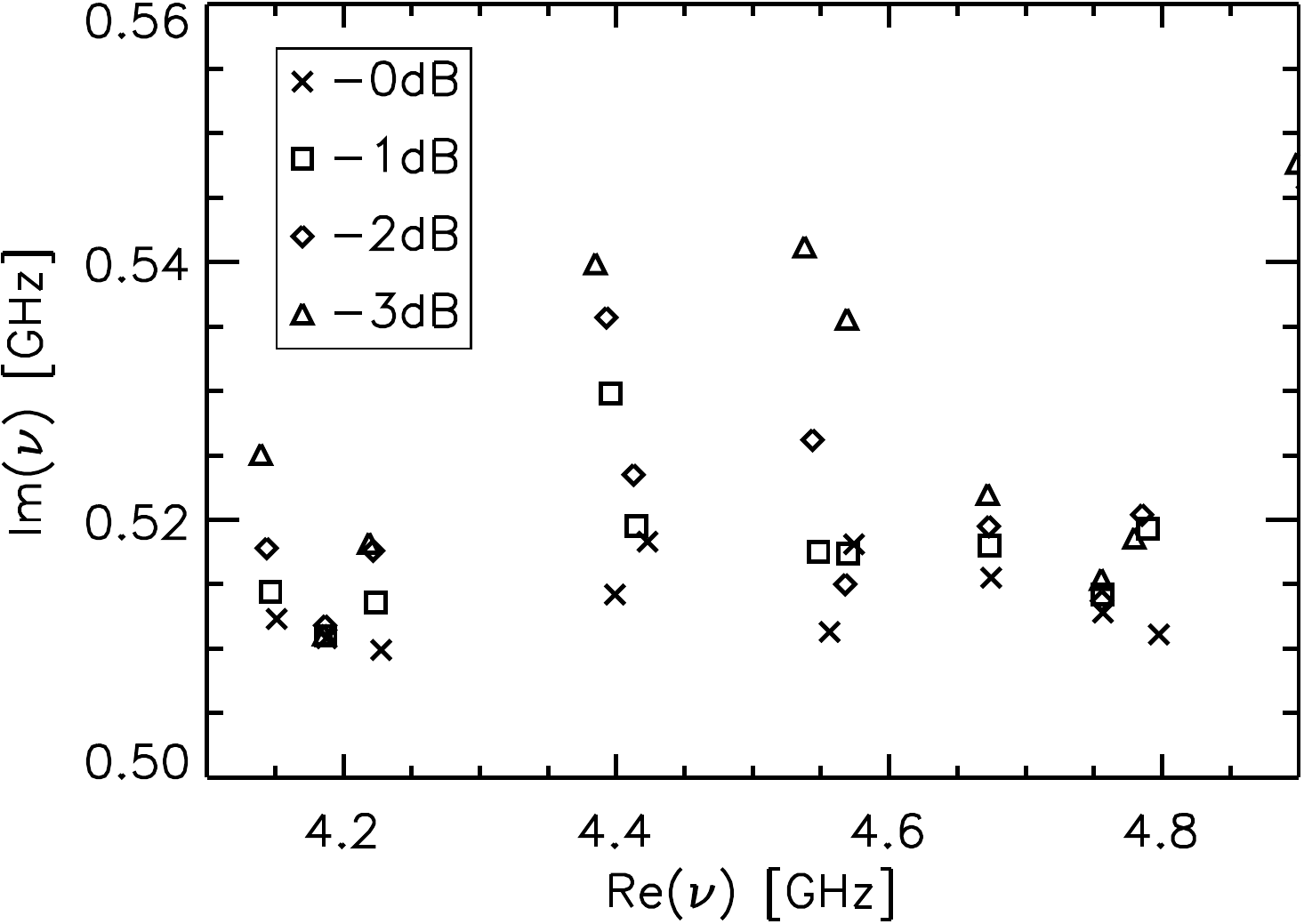}
  \caption{\label{fig:attenuation}
  Eigenfrequency spectra of the tetrahedral graph (see the inset of Fig.~\ref{fig:lengthspectrum}) in the complex plane for small attenuations $\alpha$ (0-3\,dB) in one of the bonds.}
\end{figure}

The quantum-mechanical spectrum of a system can be expressed in terms of its classical periodic orbits via the Gutzwiller trace formula \cite{gut71}. The standard derivation relies on a stationary phase approximation of the Feynman path integral, therefore, the formula is applicable only in the semiclassical limit. In this respect, quantum graphs are much simpler. The identification and classification of the periodic orbits are straightforward making them an ideal model system to study the relation between its classical and its quantum-mechanical properties. Details can be found in the paper by Kottos and Smilansky \cite{kot99a}. From an expansion of the spectral determinant they obtained a periodic orbit expansion of the spectrum, which, for graphs with TRS and von Neumann boundary conditions, read
\begin{equation}\label{eq:period}
  \rho(k)=\frac{L}{\pi}+\frac{1}{\pi}\sum\limits_{p,r}l_pA_p^r\cos{rkl_p}\,.
\end{equation}
$\rho(k)$ is the spectral density of the eigen-wavenumbers $k_n$, as is indicated by the notation, not of the eigenenergies $E_n=k_n^2$. In billiard systems, the wave number $k$ is a much more convenient quantity to look at than the energy. This is, in particular, true for the periodic orbit expansion where $k$
enters but not $E$. For one-dimensional systems, such as graphs, there is the further advantage that, with $k$ as the variable, the mean density of states is constant, which is a prerequisite for the study, e.\,g., of level-spacing distributions. The usually needed unfolding of the energy axis to a constant mean density is, thus, dispensable with $k$ as the variable.

Equation (\ref{eq:period}) bears a striking similarity to the Gutzwiller trace formula but is exact in contrast to the latter one. The first term constitutes the smooth part of the spectrum. It is proportional to the total length $L$ of the graph, reflecting the fact that graphs are one-dimensional systems (in Ref.~\cite{kot99a}, this term is written as ${\cal L}/(2\pi)$, where $\cal{L}$ corresponds to twice the total length). The second term, the fluctuating part of the spectrum, is given by a sum over all primitive periodic orbits $p$ and their repetitions $r$. $l_p$ is the length of the primitive orbit, and $A_p$ is a stability factor given by
\begin{equation}
A_p= \left(\frac{1}{3}\right)^{\mu_p}  \left(\frac{2}{3}\right)^{\nu_p}\,,
\end{equation}
where $\mu_p$ is the number of vertices in the orbit at which reflection occurs and $\nu_p$ is the number of vertices where transmission occurs. This is an immediate consequence of the specific form (\ref{eq:scatt}) of the scattering matrix of the T junction. With an attenuator in one of the bonds, this expression is modified to
\begin{equation}
A_p= a^{n_p}\left(\frac{1}{3}\right)^{\mu_p}  \left(\frac{2}{3}\right)^{\nu_p},
\end{equation}
where $n_p$ is the number of passages of orbit $p$ through the absorber. The periodic orbit expansion (\ref{eq:period}) now may be decomposed as
\begin{equation}\label{eq:absperiod}
  \rho(k)=\frac{L}{\pi}+\rho_0(k)+a\rho_1(k)+a^2\rho_2(k)+\cdots,
\end{equation}
where each $\rho_n(k)$ contains only the contribution of those orbits passing the attenuation $n$ times. The attenuation factor $a$ is related to the attenuation of the attenuator $\alpha$ given in decibel (dB) via $a=10^{-\alpha/(10\cdot 2)}$. The factor 1/2 in the exponent arises from the fact that the attenuation refers to the energy, whereas, $a$ is an amplitude. It is, hence, expected that the amplitude factor of $\rho_n(k)$ as a function of $\alpha$ decays with $10^{-n\alpha/20}$.

\begin{figure}
  \includegraphics[width=.95\columnwidth]{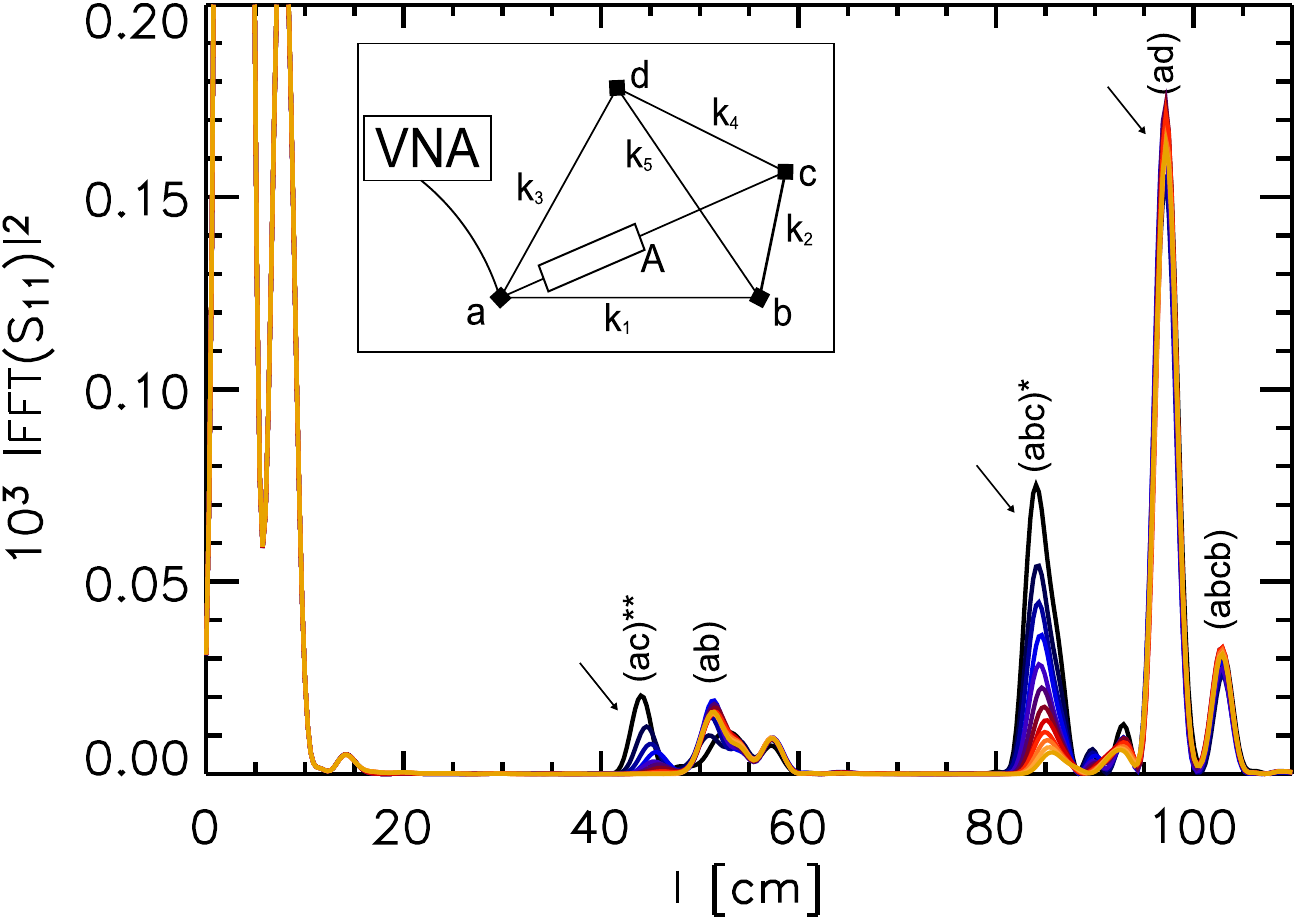}\\
  \caption{\label{fig:lengthspectrum}
  Length spectrum $|\textrm{FFT}(S_{11})|^2$ of the tetrahedral graph with attenuation in one of the bonds for different attenuations (0-11\,dB, black to yellow, respectively). The height dependences of the three peaks marked by arrows are shown in Fig.~\ref{fig:palpha}. The inset shows the configuration of the used microwave graph. The optical lengths of the cables, including the contribution of the T junctions, are $A=21.0$, $k_1=21.9$, $k_2=36.3$, $k_3=48.8$, $k_4=54.7$, and $k_5=84.9$\,cm.}
\end{figure}

The test of this expectation was one of the objects of this paper. Measurements were performed with the graph shown in Fig.~\ref{fig:setup}(b) for the time-reversal invariant graph. Figure~\ref{fig:attenuation} shows eigenfrequency spectra as a function of attenuation. In contrast to the measurements with the 50\,$\Omega$ terminator, now the eigenvalues acquire imaginary parts proportional to the attenuation coefficient $\alpha$, but the real parts are only marginally influenced by the attenuator. This is why these measurements were not suited as a test for the Poli \emph{et\,al.}\ distributions presented in Sec.~\ref{sec:poli}. We would like to emphasize that the attenuator conforming to the 50\,$\Omega$ technology, too, actually does not change the phase acquired along the bond, thus, leaving the phase difference for the two vortices at the bond ends equal. Therefore, the phase condition for the resonances stays the same, which probably is the explanation for the fact that their real parts are only weakly influenced by the attenuation, whereas, an attached 50\,$\Omega$ terminator changes the conditions at the neighboring vertices leading to a change in the real parts as well. In both cases, we do have local absorption, but it enters differently into the scattering problem. The situation is analogous to chaotic billiards, where wall absorption induces an imaginary part only to the resonances, whereas, open channels change both imaginary {\it and} real parts.

From the measurements of the $S_{11}$ reflection signal for 12 different attenuations $\alpha$ ranging from 0 to 11\,dB, we obtained the periodic orbit spectrum by applying a fast Fourier transformation (FFT) to the spectra. This length spectrum is shown in Fig.~\ref{fig:lengthspectrum}. Almost all observed peaks can be identified with periodic orbits on the graph. Orbits passing through the attenuator once or twice are marked with one or two asterisks, respectively. The peaks related to orbits not passing the attenuator are constant, whereas, the peaks related to orbits passing the attenuator are decreasing with attenuation strength $\alpha$. Additionally, we observe a small shift in orbit length, which is caused by the fact that, with increasing attenuation strength, the electrical length of the attenuator is slightly increasing. This is, as well, probably the reason why the real parts of the resonances in Fig.~\ref{fig:attenuation} are changing.

\begin{figure}
  \includegraphics[width=.95\columnwidth]{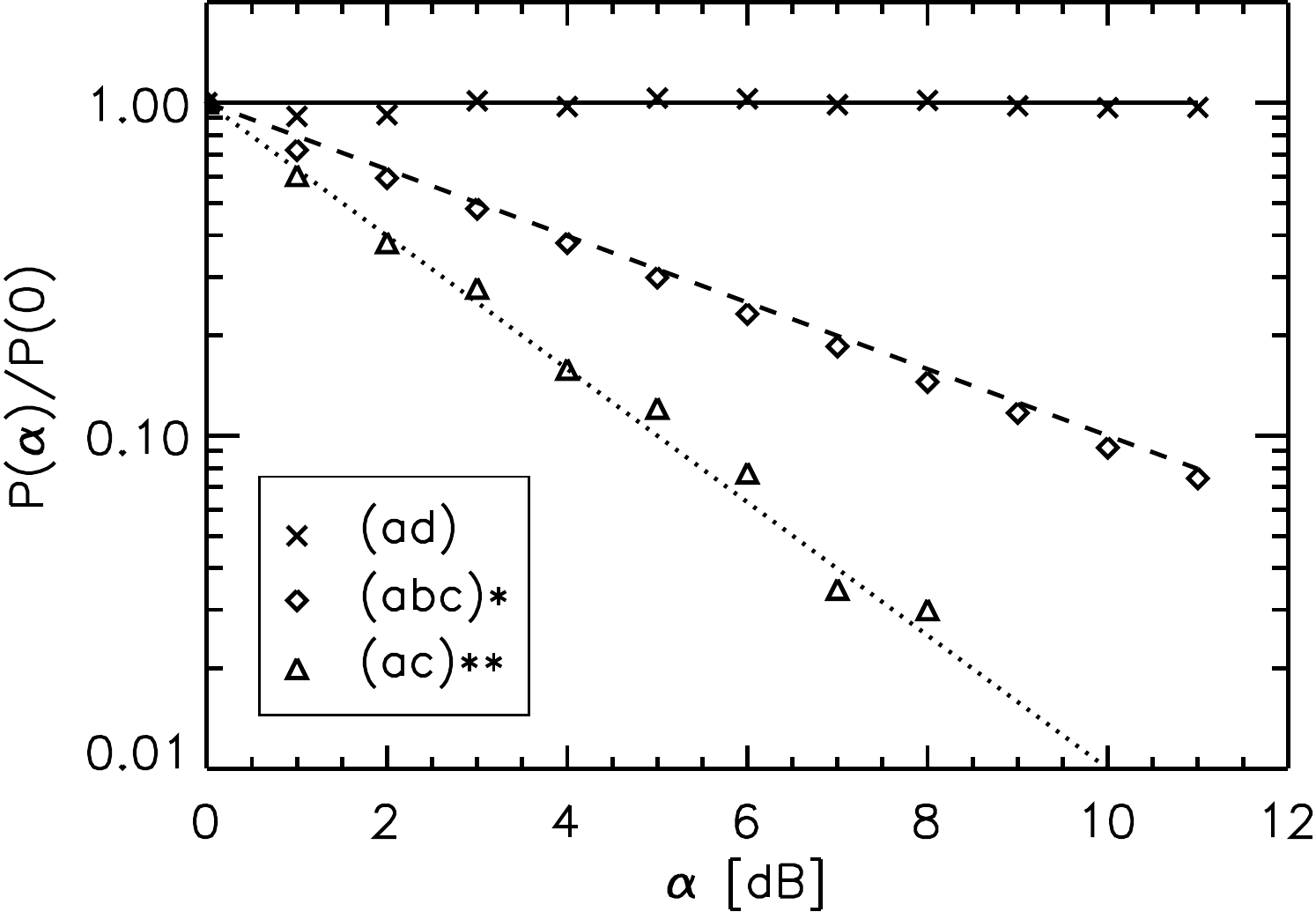}
  \caption{\label{fig:palpha}
  Amplitude of three orbits in dependence of the attenuation: one avoiding the attenuator ($\times$), one passing it once ($\diamond$), and another one passing it twice ($\triangle$). The slopes of the straight lines reflect the expectations from Eq.~(\ref{eq:absperiod}).}
\end{figure}

To investigate the decrease in the peaks with attenuation strength in a quantitative way, in Fig.~\ref{fig:palpha}, the heights of three peaks are shown in dependence of the attenuation. One corresponds to an orbit avoiding the attenuator, one to an orbit passing the attenuator once, and another one passing the attenuator twice. For the orbit avoiding the attenuator, the amplitude is found to be nearly independent of the attenuation, whereas, for the orbits passing the attenuator once and twice, the expected exponential decay of the amplitude with $\alpha$ is observed. Analogous behaviors were found for all identified orbits, thus, illustrating the periodic orbit decomposition of Eq.~(\ref{eq:absperiod}).

\section{Conclusions}

We have shown that microwave networks can not only reproduce the results for the nearest-neighbor distance distribution of closed quantum graphs \cite{hul04}, but also may serve as a means of investigating the influence of open channels. As an example, we could verify the level-spacing distribution proposed by Poli \emph{et\,al.}~\cite{pol12} for systems with attached open channels both for the GOE and for the GUE. Furthermore, we were able to investigate the influence of local attenuation in a bond on the periodic orbit spectrum of quantum graphs. In this context, we were able to show that attenuation on one bond of the graph directly influences the amplitudes of periodic orbits in the length spectrum that includes the microwave attenuator.

\begin{acknowledgments}
This work was supported by the Deutsche Forschungsgemeinschaft via the Forschergruppe 760 ``Scattering systems with complex dynamics.'' U.~Smilansky and O.~Hul are thanked for fruitful discussions.
\end{acknowledgments}


\begin{thebibliography}{30}%
\makeatletter
\providecommand \@ifxundefined [1]{%
 \@ifx{#1\undefined}
}%
\providecommand \@ifnum [1]{%
 \ifnum #1\expandafter \@firstoftwo
 \else \expandafter \@secondoftwo
 \fi
}%
\providecommand \@ifx [1]{%
 \ifx #1\expandafter \@firstoftwo
 \else \expandafter \@secondoftwo
 \fi
}%
\providecommand \natexlab [1]{#1}%
\providecommand \enquote  [1]{``#1''}%
\providecommand \bibnamefont  [1]{#1}%
\providecommand \bibfnamefont [1]{#1}%
\providecommand \citenamefont [1]{#1}%
\providecommand \href@noop [0]{\@secondoftwo}%
\providecommand \href [0]{\begingroup \@sanitize@url \@href}%
\providecommand \@href[1]{\@@startlink{#1}\@@href}%
\providecommand \@@href[1]{\endgroup#1\@@endlink}%
\providecommand \@sanitize@url [0]{\catcode `\\12\catcode `\$12\catcode
  `\&12\catcode `\#12\catcode `\^12\catcode `\_12\catcode `\%12\relax}%
\providecommand \@@startlink[1]{}%
\providecommand \@@endlink[0]{}%
\providecommand \url  [0]{\begingroup\@sanitize@url \@url }%
\providecommand \@url [1]{\endgroup\@href {#1}{\urlprefix }}%
\providecommand \urlprefix  [0]{URL }%
\providecommand \Eprint [0]{\href }%
\@ifxundefined \urlstyle {%
  \providecommand \doi  [0]{\begingroup \@sanitize@url \@doi}%
  \providecommand \@doi [1]{\endgroup \@@startlink {\doibase
  #1}doi:\discretionary {}{}{}#1\@@endlink }%
}{%
  \providecommand \doi  [0]{doi:\discretionary{}{}{}\begingroup
  \urlstyle{rm}\Url }%
}%
\providecommand \doibase [0]{http://dx.doi.org/}%
\providecommand \Doi [0]{\begingroup \@sanitize@url \@Doi }%
\providecommand \@Doi  [1]{\endgroup\@@startlink{\doibase#1}\@@Doi}%
\providecommand \@@Doi [1]{#1\@@endlink}%
\providecommand \selectlanguage [0]{\@gobble}%
\providecommand \bibinfo  [0]{\@secondoftwo}%
\providecommand \bibfield  [0]{\@secondoftwo}%
\providecommand \translation [1]{[#1]}%
\providecommand \BibitemOpen [0]{}%
\providecommand \bibitemStop [0]{}%
\providecommand \bibitemNoStop [0]{.\EOS\space}%
\providecommand \EOS [0]{\spacefactor3000\relax}%
\providecommand \BibitemShut  [1]{\csname bibitem#1\endcsname}%
\bibitem [{\citenamefont {Gutzwiller}(1990)}]{gut90}%
  \BibitemOpen
  \bibfield  {author} {\bibinfo {author} {\bibfnamefont {M.~C.}\ \bibnamefont
  {Gutzwiller}},\ }\href@noop {} {\emph {\bibinfo {title} {Chaos in Classical
  and Quantum Mechanics}}},\ Interdisciplinary Applied Mathematics, Vol. 1\
  (\bibinfo  {publisher} {Springer},\ \bibinfo {address} {New York},\ \bibinfo
  {year} {1990})\BibitemShut {NoStop}%
\bibitem [{\citenamefont {Kottos}\ and\ \citenamefont
  {Smilansky}(1999)}]{kot99a}%
  \BibitemOpen
  \bibfield  {author} {\bibinfo {author} {\bibfnamefont {T.}~\bibnamefont
  {Kottos}}\ and\ \bibinfo {author} {\bibfnamefont {U.}~\bibnamefont
  {Smilansky}},\ }\Doi {10.1006/aphy.1999.5904} {\bibfield  {journal} {\bibinfo
   {journal} {Ann. Phys. (N.Y.)},\ }\textbf {\bibinfo {volume} {274}},\
  \bibinfo {pages} {76} (\bibinfo {year} {1999})}\BibitemShut {NoStop}%
\bibitem [{\citenamefont {Bohigas}\ \emph {et~al.}(1984)\citenamefont
  {Bohigas}, \citenamefont {Giannoni},\ and\ \citenamefont {Schmit}}]{boh84b}%
  \BibitemOpen
  \bibfield  {author} {\bibinfo {author} {\bibfnamefont {O.}~\bibnamefont
  {Bohigas}}, \bibinfo {author} {\bibfnamefont {M.~J.}\ \bibnamefont
  {Giannoni}}, \ and\ \bibinfo {author} {\bibfnamefont {C.}~\bibnamefont
  {Schmit}},\ }\href@noop {} {\bibfield  {journal} {\bibinfo  {journal} {Phys.
  Rev. Lett.},\ }\textbf {\bibinfo {volume} {52}},\ \bibinfo {pages} {1}
  (\bibinfo {year} {1984})}\BibitemShut {NoStop}%
\bibitem [{\citenamefont {Casati}\ \emph {et~al.}(1980)\citenamefont {Casati},
  \citenamefont {Valz-Gris},\ and\ \citenamefont {Guarnieri}}]{cas80}%
  \BibitemOpen
  \bibfield  {author} {\bibinfo {author} {\bibfnamefont {G.}~\bibnamefont
  {Casati}}, \bibinfo {author} {\bibfnamefont {F.}~\bibnamefont {Valz-Gris}}, \
  and\ \bibinfo {author} {\bibfnamefont {I.}~\bibnamefont {Guarnieri}},\
  }\href@noop {} {\bibfield  {journal} {\bibinfo  {journal} {Lett. Nuov.
  Cim.},\ }\textbf {\bibinfo {volume} {28}},\ \bibinfo {pages} {279} (\bibinfo
  {year} {1980})}\BibitemShut {NoStop}%
\bibitem [{\citenamefont {Gnutzmann}\ and\ \citenamefont
  {Altland}(2004)}]{gnu04b}%
  \BibitemOpen
  \bibfield  {author} {\bibinfo {author} {\bibfnamefont {S.}~\bibnamefont
  {Gnutzmann}}\ and\ \bibinfo {author} {\bibfnamefont {A.}~\bibnamefont
  {Altland}},\ }\Doi {10.1103/PhysRevLett.93.194101} {\bibfield  {journal}
  {\bibinfo  {journal} {Phys. Rev. Lett.},\ }\textbf {\bibinfo {volume} {93}},\
  \bibinfo {pages} {194101} (\bibinfo {year} {2004})}\BibitemShut {NoStop}%
\bibitem [{\citenamefont {Pluhar}\ and\ \citenamefont
  {Weidenm\"{u}ller}(2013)}]{arXplu13}%
  \BibitemOpen
  \bibfield  {author} {\bibinfo {author} {\bibfnamefont {Z.}~\bibnamefont
  {Pluhar}}\ and\ \bibinfo {author} {\bibfnamefont {H.~A.}\ \bibnamefont
  {Weidenm\"{u}ller}},\ }\href@noop {} {\enquote {\bibinfo {title} {Universal
  quantum graphs},}\ }\bibinfo {howpublished} {Preprint} (\bibinfo {year}
  {2013}),\ \bibinfo {note} {arXiv:1312.2349}\BibitemShut {NoStop}%
\bibitem [{\citenamefont {Hul}\ \emph {et~al.}(2004)\citenamefont {Hul},
  \citenamefont {Bauch}, \citenamefont {Pako\~{n}ski}, \citenamefont
  {Savytskyy}, \citenamefont {\.{Z}yczkowski},\ and\ \citenamefont
  {Sirko}}]{hul04}%
  \BibitemOpen
  \bibfield  {author} {\bibinfo {author} {\bibfnamefont {O.}~\bibnamefont
  {Hul}}, \bibinfo {author} {\bibfnamefont {S.}~\bibnamefont {Bauch}}, \bibinfo
  {author} {\bibfnamefont {P.}~\bibnamefont {Pako\~{n}ski}}, \bibinfo {author}
  {\bibfnamefont {N.}~\bibnamefont {Savytskyy}}, \bibinfo {author}
  {\bibfnamefont {K.}~\bibnamefont {\.{Z}yczkowski}}, \ and\ \bibinfo {author}
  {\bibfnamefont {L.}~\bibnamefont {Sirko}},\ }\href@noop {} {\bibfield
  {journal} {\bibinfo  {journal} {Phys. Rev. E},\ }\textbf {\bibinfo {volume}
  {69}},\ \bibinfo {pages} {056205} (\bibinfo {year} {2004})}\BibitemShut
  {NoStop}%
\bibitem [{\citenamefont {St\"{o}ckmann}\ and\ \citenamefont {{\v
  S}eba}(1998)}]{stoe98}%
  \BibitemOpen
  \bibfield  {author} {\bibinfo {author} {\bibfnamefont {H.-J.}\ \bibnamefont
  {St\"{o}ckmann}}\ and\ \bibinfo {author} {\bibfnamefont {P.}~\bibnamefont
  {{\v S}eba}},\ }\href@noop {} {\bibfield  {journal} {\bibinfo  {journal} {J.
  Phys. A},\ }\textbf {\bibinfo {volume} {31}},\ \bibinfo {pages} {3439}
  (\bibinfo {year} {1998})}\BibitemShut {NoStop}%
\bibitem [{\citenamefont {Haake}(2001)}]{haa01b}%
  \BibitemOpen
  \bibfield  {author} {\bibinfo {author} {\bibfnamefont {F.}~\bibnamefont
  {Haake}},\ }\href@noop {} {\emph {\bibinfo {title} {Quantum Signatures of
  Chaos. 2nd edition}}}\ (\bibinfo  {publisher} {Springer},\ \bibinfo {address}
  {Berlin},\ \bibinfo {year} {2001})\BibitemShut {NoStop}%
\bibitem [{\citenamefont {Poli}\ \emph {et~al.}(2012)\citenamefont {Poli},
  \citenamefont {Luna-Acosta},\ and\ \citenamefont {St\"{o}ckmann}}]{pol12}%
  \BibitemOpen
  \bibfield  {author} {\bibinfo {author} {\bibfnamefont {C.}~\bibnamefont
  {Poli}}, \bibinfo {author} {\bibfnamefont {G.~A.}\ \bibnamefont
  {Luna-Acosta}}, \ and\ \bibinfo {author} {\bibfnamefont {H.-J.}\ \bibnamefont
  {St\"{o}ckmann}},\ }\Doi {10.1103/PhysRevLett.108.174101} {\bibfield
  {journal} {\bibinfo  {journal} {Phys. Rev. Lett.},\ }\textbf {\bibinfo
  {volume} {108}},\ \bibinfo {pages} {174101} (\bibinfo {year}
  {2012})}\BibitemShut {NoStop}%
\bibitem [{\citenamefont {Hul}\ \emph {et~al.}(2005)\citenamefont {Hul},
  \citenamefont {Tymoshchuk}, \citenamefont {Bauch}, \citenamefont {Koch},\
  and\ \citenamefont {Sirko}}]{hul05a}%
  \BibitemOpen
  \bibfield  {author} {\bibinfo {author} {\bibfnamefont {O.}~\bibnamefont
  {Hul}}, \bibinfo {author} {\bibfnamefont {O.}~\bibnamefont {Tymoshchuk}},
  \bibinfo {author} {\bibfnamefont {S.}~\bibnamefont {Bauch}}, \bibinfo
  {author} {\bibfnamefont {P.}~\bibnamefont {Koch}}, \ and\ \bibinfo {author}
  {\bibfnamefont {L.}~\bibnamefont {Sirko}},\ }\Doi
  {10.1088/0305-4470/38/49/003} {\bibfield  {journal} {\bibinfo  {journal} {J.
  Phys. A},\ }\textbf {\bibinfo {volume} {38}},\ \bibinfo {pages} {10489}
  (\bibinfo {year} {2005})}\BibitemShut {NoStop}%
\bibitem [{\citenamefont {{\L}awniczak}\ \emph {et~al.}(2010)\citenamefont
  {{\L}awniczak}, \citenamefont {Bauch}, \citenamefont {Hul},\ and\
  \citenamefont {Sirko}}]{law10}%
  \BibitemOpen
  \bibfield  {author} {\bibinfo {author} {\bibfnamefont {M.}~\bibnamefont
  {{\L}awniczak}}, \bibinfo {author} {\bibfnamefont {S.}~\bibnamefont {Bauch}},
  \bibinfo {author} {\bibfnamefont {O.}~\bibnamefont {Hul}}, \ and\ \bibinfo
  {author} {\bibfnamefont {L.}~\bibnamefont {Sirko}},\ }\Doi
  {10.1103/PhysRevE.81.046204} {\bibfield  {journal} {\bibinfo  {journal}
  {Phys. Rev. E},\ }\textbf {\bibinfo {volume} {81}},\ \bibinfo {pages}
  {046204} (\bibinfo {year} {2010})}\BibitemShut {NoStop}%
\bibitem [{\citenamefont {Hul}\ \emph {et~al.}(2012)\citenamefont {Hul},
  \citenamefont {{\L}awniczak}, \citenamefont {Bauch}, \citenamefont {Sawicki},
  \citenamefont {Ku\'s},\ and\ \citenamefont {Sirko}}]{hul12}%
  \BibitemOpen
  \bibfield  {author} {\bibinfo {author} {\bibfnamefont {O.}~\bibnamefont
  {Hul}}, \bibinfo {author} {\bibfnamefont {M.}~\bibnamefont {{\L}awniczak}},
  \bibinfo {author} {\bibfnamefont {S.}~\bibnamefont {Bauch}}, \bibinfo
  {author} {\bibfnamefont {A.}~\bibnamefont {Sawicki}}, \bibinfo {author}
  {\bibfnamefont {M.}~\bibnamefont {Ku\'s}}, \ and\ \bibinfo {author}
  {\bibfnamefont {L.}~\bibnamefont {Sirko}},\ }\href@noop {} {\bibfield
  {journal} {\bibinfo  {journal} {Phys. Rev. Lett.},\ }\textbf {\bibinfo
  {volume} {109}},\ \bibinfo {pages} {040402} (\bibinfo {year}
  {2012})}\BibitemShut {NoStop}%
\bibitem [{\citenamefont {Kottos}\ and\ \citenamefont
  {Smilansky}(2003)}]{kot03}%
  \BibitemOpen
  \bibfield  {author} {\bibinfo {author} {\bibfnamefont {T.}~\bibnamefont
  {Kottos}}\ and\ \bibinfo {author} {\bibfnamefont {U.}~\bibnamefont
  {Smilansky}},\ }\Doi {10.1088/0305-4470/36/12/337} {\bibfield  {journal}
  {\bibinfo  {journal} {J. Phys. A},\ }\textbf {\bibinfo {volume} {36}},\
  \bibinfo {pages} {3501} (\bibinfo {year} {2003})}\BibitemShut {NoStop}%
\bibitem [{\citenamefont {Mart\'{\i}nez-Mendoza}\ \emph
  {et~al.}(2013)\citenamefont {Mart\'{\i}nez-Mendoza}, \citenamefont
  {Alcazar-L\'{o}pez},\ and\ \citenamefont {M\'{e}ndez-Berm\'{u}dez}}]{mar13}%
  \BibitemOpen
  \bibfield  {author} {\bibinfo {author} {\bibfnamefont {A.~J.}\ \bibnamefont
  {Mart\'{\i}nez-Mendoza}}, \bibinfo {author} {\bibfnamefont {A.}~\bibnamefont
  {Alcazar-L\'{o}pez}}, \ and\ \bibinfo {author} {\bibfnamefont {J.~A.}\
  \bibnamefont {M\'{e}ndez-Berm\'{u}dez}},\ }\Doi {10.1103/PhysRevE.88.012126}
  {\bibfield  {journal} {\bibinfo  {journal} {Phys. Rev. E},\ }\textbf
  {\bibinfo {volume} {88}},\ \bibinfo {pages} {012126} (\bibinfo {year}
  {2013})}\BibitemShut {NoStop}%
\bibitem [{\citenamefont {Gnutzmann}\ \emph {et~al.}(2013)\citenamefont
  {Gnutzmann}, \citenamefont {Schanz},\ and\ \citenamefont
  {Smilansky}}]{gnu13}%
  \BibitemOpen
  \bibfield  {author} {\bibinfo {author} {\bibfnamefont {S.}~\bibnamefont
  {Gnutzmann}}, \bibinfo {author} {\bibfnamefont {H.}~\bibnamefont {Schanz}}, \
  and\ \bibinfo {author} {\bibfnamefont {U.}~\bibnamefont {Smilansky}},\ }\Doi
  {10.1103/PhysRevLett.110.094101} {\bibfield  {journal} {\bibinfo  {journal}
  {Phys. Rev. Lett.},\ }\textbf {\bibinfo {volume} {110}},\ \bibinfo {pages}
  {094101} (\bibinfo {year} {2013})}\BibitemShut {NoStop}%
\bibitem [{\citenamefont {Gutkin}\ and\ \citenamefont {Osipov}(2013)}]{arXgut}%
  \BibitemOpen
  \bibfield  {author} {\bibinfo {author} {\bibfnamefont {B.}~\bibnamefont
  {Gutkin}}\ and\ \bibinfo {author} {\bibfnamefont {V.~A.}\ \bibnamefont
  {Osipov}},\ }\href@noop {} {\enquote {\bibinfo {title} {Universality in
  spectral statistics of ``open'' quantum graphs.}}\ } (\bibinfo {year}
  {2013}),\ \bibinfo {note} {arXiv:1308.2356}\BibitemShut {NoStop}%
\bibitem [{\citenamefont {Pluha{\v r}}\ and\ \citenamefont
  {Weidenm\"{u}ller}(2013){\natexlab{a}}}]{plu13a}%
  \BibitemOpen
  \bibfield  {author} {\bibinfo {author} {\bibfnamefont {Z.}~\bibnamefont
  {Pluha{\v r}}}\ and\ \bibinfo {author} {\bibfnamefont {H.~A.}\ \bibnamefont
  {Weidenm\"{u}ller}},\ }\Doi {10.1103/PhysRevLett.110.034101} {\bibfield
  {journal} {\bibinfo  {journal} {Phys. Rev. Lett.},\ }\textbf {\bibinfo
  {volume} {110}},\ \bibinfo {pages} {034101} (\bibinfo {year}
  {2013}{\natexlab{a}})}\BibitemShut {NoStop}%
\bibitem [{\citenamefont {Pluha{\v r}}\ and\ \citenamefont
  {Weidenm\"{u}ller}(2013){\natexlab{b}}}]{plu13b}%
  \BibitemOpen
  \bibfield  {author} {\bibinfo {author} {\bibfnamefont {Z.}~\bibnamefont
  {Pluha{\v r}}}\ and\ \bibinfo {author} {\bibfnamefont {H.~A.}\ \bibnamefont
  {Weidenm\"{u}ller}},\ }\Doi {10.1103/PhysRevE.88.022902} {\bibfield
  {journal} {\bibinfo  {journal} {Phys. Rev. E},\ }\textbf {\bibinfo {volume}
  {88}},\ \bibinfo {pages} {022902} (\bibinfo {year}
  {2013}{\natexlab{b}})}\BibitemShut {NoStop}%
\bibitem [{\citenamefont {Jackson}(1998)}]{jac98}%
  \BibitemOpen
  \bibfield  {author} {\bibinfo {author} {\bibfnamefont {J.~D.}\ \bibnamefont
  {Jackson}},\ }\href@noop {} {\emph {\bibinfo {title} {Classical Electrodynamics, 3rd ed.}}}\ (\bibinfo  {publisher} {John Wiley \& Sons},\ \bibinfo
  {address} {New York},\ \bibinfo {year} {1998})\BibitemShut {NoStop}%
\bibitem [{\citenamefont {Beenakker}(1997)}]{bee97}%
  \BibitemOpen
  \bibfield  {author} {\bibinfo {author} {\bibfnamefont {C.~W.~J.}\
  \bibnamefont {Beenakker}},\ }\href@noop {} {\bibfield  {journal} {\bibinfo
  {journal} {Rev. Mod. Phys.},\ }\textbf {\bibinfo {volume} {69}},\ \bibinfo
  {pages} {731} (\bibinfo {year} {1997})}\BibitemShut {NoStop}%
\bibitem [{\citenamefont {St\"{o}ckmann}(1999)}]{stoe99}%
  \BibitemOpen
  \bibfield  {author} {\bibinfo {author} {\bibfnamefont {H.-J.}\ \bibnamefont
  {St\"{o}ckmann}},\ }\href@noop {} {\emph {\bibinfo {title} {Quantum Chaos -
  An Introduction}}}\ (\bibinfo  {publisher} {University Press},\ \bibinfo
  {address} {Cambridge},\ \bibinfo {year} {1999})\BibitemShut {NoStop}%
\bibitem [{\citenamefont {Kahn}(1962)}]{kah62}%
  \BibitemOpen
  \bibfield  {author} {\bibinfo {author} {\bibfnamefont {P.~B.}\ \bibnamefont
  {Kahn}},\ }\href@noop {} {\bibfield  {journal} {\bibinfo  {journal} {Nuclear
  Physics},\ }\textbf {\bibinfo {volume} {41}},\ \bibinfo {pages} {159}
  (\bibinfo {year} {1962})}\BibitemShut {NoStop}%
\bibitem [{\citenamefont {Berry}\ and\ \citenamefont {Robnik}(1984)}]{ber84b}%
  \BibitemOpen
  \bibfield  {author} {\bibinfo {author} {\bibfnamefont {M.~V.}\ \bibnamefont
  {Berry}}\ and\ \bibinfo {author} {\bibfnamefont {M.}~\bibnamefont {Robnik}},\
  }\href@noop {} {\bibfield  {journal} {\bibinfo  {journal} {J. Phys. A},\
  }\textbf {\bibinfo {volume} {17}},\ \bibinfo {pages} {2413} (\bibinfo {year}
  {1984})}\BibitemShut {NoStop}%
\bibitem [{\citenamefont {Barra}\ and\ \citenamefont {Gaspard}(2000)}]{bar00b}%
  \BibitemOpen
  \bibfield  {author} {\bibinfo {author} {\bibfnamefont {F.}~\bibnamefont
  {Barra}}\ and\ \bibinfo {author} {\bibfnamefont {P.}~\bibnamefont
  {Gaspard}},\ }\Doi {10.1023/A:1026495012522} {\bibfield  {journal} {\bibinfo
  {journal} {J. Stat. Phys.},\ }\textbf {\bibinfo {volume} {101}},\ \bibinfo
  {pages} {283} (\bibinfo {year} {2000})}\BibitemShut {NoStop}%
\bibitem [{\citenamefont {Mehta}(1991)}]{meh91}%
  \BibitemOpen
  \bibfield  {author} {\bibinfo {author} {\bibfnamefont {M.~L.}\ \bibnamefont
  {Mehta}},\ }\href@noop {} {\emph {\bibinfo {title} {Random Matrices. 2nd
  edition}}}\ (\bibinfo  {publisher} {Academic Press},\ \bibinfo {address} {San
  Diego},\ \bibinfo {year} {1991})\BibitemShut {NoStop}%
\bibitem [{\citenamefont {Barth\'elemy}\ \emph {et~al.}(2005)\citenamefont
  {Barth\'elemy}, \citenamefont {Legrand},\ and\ \citenamefont
  {Mortessagne}}]{bar05a}%
  \BibitemOpen
  \bibfield  {author} {\bibinfo {author} {\bibfnamefont {J.}~\bibnamefont
  {Barth\'elemy}}, \bibinfo {author} {\bibfnamefont {O.}~\bibnamefont
  {Legrand}}, \ and\ \bibinfo {author} {\bibfnamefont {F.}~\bibnamefont
  {Mortessagne}},\ }\href@noop {} {\bibfield  {journal} {\bibinfo  {journal}
  {Phys. Rev. E},\ }\textbf {\bibinfo {volume} {71}},\ \bibinfo {pages}
  {016205} (\bibinfo {year} {2005})}\BibitemShut {NoStop}%
\bibitem [{\citenamefont {Verbaarschot}\ \emph {et~al.}(1985)\citenamefont
  {Verbaarschot}, \citenamefont {Weidenm\"{u}ller},\ and\ \citenamefont
  {Zirnbauer}}]{ver85a}%
  \BibitemOpen
  \bibfield  {author} {\bibinfo {author} {\bibfnamefont {J.~J.~M.}\
  \bibnamefont {Verbaarschot}}, \bibinfo {author} {\bibfnamefont {H.~A.}\
  \bibnamefont {Weidenm\"{u}ller}}, \ and\ \bibinfo {author} {\bibfnamefont
  {M.~R.}\ \bibnamefont {Zirnbauer}},\ }\href@noop {} {\bibfield  {journal}
  {\bibinfo  {journal} {Phys. Rep.},\ }\textbf {\bibinfo {volume} {129}},\
  \bibinfo {pages} {367} (\bibinfo {year} {1985})}\BibitemShut {NoStop}%
\bibitem [{\citenamefont {K\"{o}ber}\ \emph {et~al.}(2010)\citenamefont
  {K\"{o}ber}, \citenamefont {Kuhl}, \citenamefont {St\"{o}ckmann},
  \citenamefont {Gorin}, \citenamefont {Savin},\ and\ \citenamefont
  {Seligman}}]{koeb10}%
  \BibitemOpen
  \bibfield  {author} {\bibinfo {author} {\bibfnamefont {B.}~\bibnamefont
  {K\"{o}ber}}, \bibinfo {author} {\bibfnamefont {U.}~\bibnamefont {Kuhl}},
  \bibinfo {author} {\bibfnamefont {H.-J.}\ \bibnamefont {St\"{o}ckmann}},
  \bibinfo {author} {\bibfnamefont {T.}~\bibnamefont {Gorin}}, \bibinfo
  {author} {\bibfnamefont {D.~V.}\ \bibnamefont {Savin}}, \ and\ \bibinfo
  {author} {\bibfnamefont {T.~H.}\ \bibnamefont {Seligman}},\ }\Doi
  {10.1103/PhysRevE.82.036207} {\bibfield  {journal} {\bibinfo  {journal}
  {Phys. Rev. E},\ }\textbf {\bibinfo {volume} {82}},\ \bibinfo {pages}
  {036207} (\bibinfo {year} {2010})}\BibitemShut {NoStop}%
\bibitem [{\citenamefont {Gutzwiller}(1971)}]{gut71}%
  \BibitemOpen
  \bibfield  {author} {\bibinfo {author} {\bibfnamefont {M.~C.}\ \bibnamefont
  {Gutzwiller}},\ }\href@noop {} {\bibfield  {journal} {\bibinfo  {journal} {J.
  Math. Phys.},\ }\textbf {\bibinfo {volume} {12}},\ \bibinfo {pages} {343}
  (\bibinfo {year} {1971})}\BibitemShut {NoStop}%
\end{thebibliography}
\end{document}